\documentclass[useAMS,usenatbib]{mn2e}
\usepackage{graphicx}
\usepackage{amssymb}

\title[Modelling rapid TeV variability of PKS 2155--304]{Modelling rapid TeV variability of PKS 2155--304}
\author[K.~Katarzy\'nski, J-P.~Lenain, A.~Zech, C.~Boisson, and H.~Sol]
       {K.~Katarzy\'nski,$^{1}$\thanks{E-mail:kat@astro.uni.torun.pl}, J-P.~Lenain$^{2}$, A.~Zech$^{2}$, C.~Boisson$^{2}$, and H.~Sol$^{2}$\\
$^{1}$Toru\'n Centre for Astronomy, Nicolaus Copernicus University, ul. Gagarina 11, PL 87-100 Toru\'n, Poland\\
$^{2}$LUTH, Observatoire de Paris, CNRS, Universit\'e Paris Diderot; 5 Place Jules Janssen, 92190 Meudon, France
}
\begin{document}

\date{Accepted 2008 July 22.  Received 2008 June 9; in original form 2008 February 1}

\pagerange{\pageref{firstpage}--\pageref{lastpage}} \pubyear{2005}

\maketitle

\label{firstpage}

\begin{abstract}
We present theoretical modelling for the very rapid TeV variability of PKS 2155--304 
observed recently by the H.E.S.S. experiment. To explain the light-curve, where at 
least five flaring events were well observed, we assume five independent components 
of a jet that are characterized by slightly different physical parameters. An additional, 
significantly larger component is used to explain the emission of the source at long time 
scales. This component dominates the emission in the X-ray range, whereas the other 
components are dominant in the TeV range. The model used for our simulation describes 
precisely the evolution of the particle energy spectrum inside each component and takes 
into account light travel time effects. We show that a relatively simple 
synchrotron self-Compton scenario may explain this very rapid variability. Moreover, 
we find that absorption of the TeV emission inside the components due to the pair 
creation process is negligible.
\end{abstract}


\section{Introduction}

Among the different models proposed to explain X-ray and gamma-ray
emission of TeV blazars one distinguishes two classes: the leptonic
and hadronic models (e.g. \citealt{Krawczynski04}). This division
depends on the assumption one makes about the particles that initially
carry the energy that is to be converted into electromagnetic emission
during the evolution of the source.  The aim of this paper is to
explain the rapid variability in the simplest possible way in order to
constrain the physical parameters of the source. Therefore, we have
decided to use a relatively simple leptonic scenario.

A very basic leptonic model that can be used to explain the high
energy emission of TeV blazars assumes a spherical source filled with
a tangled magnetic field and relativistic electrons. This source is
thought to travel with relativistic velocity at a distance of less
than 1 pc from the central black hole and is usually assumed to be
homogeneous.  The electrons inside the source produce synchrotron
emission and also up-scatter part of this emission to TeV
energies. This is the well known synchrotron self-Compton process
(SSC) that is often applied to describe X-ray and gamma-ray
observations of blazars (e.g. \citealt{Bloom96},
\citealt{Ghisellini96}, \citealt{Inoue96}, \citealt{Mastichiadis97},
\citealt{Krawczynski00}).  The simplicity of this model allows to
constrain some important physical parameters of the source
(e.g. magnetic field strength, Doppler factor) directly from observations 
(e.g. \citealt{Bednarek97}, \citealt{Tavecchio98}, \citealt{Katarzynski01}). 
On the other hand, some uncertainty remains in such estimations because
the model is too simple; it does for example not take into account
significant processes such as the particle evolution inside the source
or light crossing time effects (hereafter LCTE). Therefore it is very
important to analyse not only the emission observed at a given time
but also the evolution of the emission, especially during the periods
of activity. This may provide additional constraints for the models
and result in a more precise estimation of the physical parameters.

Several leptonic models have been proposed to explain the
time-dependent emission of TeV blazars (e.g. \citealt{Dermer97},
\citealt{Kirk98}, \citealt{Coppi99}, \citealt{Chiaberge99},
\citealt{Kusunose00}, \citealt{Kataoka00}, 
\citealt{Sokolov04}). So far, no model is able to precisely describe SSC 
emission of an inhomogeneous or even a homogeneous source. The main 
problem appears in the description of the synchrotron radiation field inside the
source that is inhomogeneous even in a spherically homogeneous slowly 
evolving source (\citealt{Gould79}). Moreover, this
radiation field depends on the particle energy distribution and the
particle energy distribution depends in turn on the radiation
field. This is certainly only true if the energy density of the
radiation field is comparable to or larger than the energy density of 
the magnetic field, but this condition is fulfilled in TeV
blazars. Finally, very rapid variability time scales are usually
assumed to be comparable to the light crossing time at the
source. This means that LCTE must be taken into account in
calculations of the inverse-Compton scattering. In other words, the
radiation field at a given time and a given position inside the source
depends not only on the local physical conditions, but also contains
photons from other parts of the source. These photons were already
created at an earlier time and have propagated to the observed
position.  We will call this process {\it internal} LCTE. This effect
is especially important for the calculation of the
inverse-Compton emission and somewhat less for the
synchrotron emission (although the radiation field has generally 
also an impact on the evolution of the electron energy spectrum). 
The internal LCTE was taken into account for the first time in the model 
proposed by \citet{Sokolov04}. However, this model assumes dominance of 
the energy density of the magnetic field over that of the radiation field.
In this case the impact of the radiation field on the electron energy 
spectrum is negligible.

A second LCTE, which we call {\it external}, is equally important for both the 
synchrotron and the inverse-Compton emission. The external observer 
receives at a given time the emission produced by different parts of the source at 
different times in the comoving frame of the source. This is due to the different travel 
times that photons from different parts of the source need to reach 
the observer. If the emission level of the source does not change,
the observer will always receive the same amount of 
radiation; if it changes on long time scales, differences in photon travel times
can be neglected. However, if there is a change of the emission level with
a duration shorter than the light crossing time of the source, the 
different travel times of the photons must be taken into account. One of the 
first attempts to account for this effect was made by \citet{Chiaberge99}. 
It should be noted that in their model they neglected internal LCTE, which is a good
enough approximation under certain conditions, as we will explain below in the description 
of the model.

Another important problem in time dependent SSC modelling is the description of 
the particle acceleration and evolution of the particle energy spectrum. 
It is widely believed that the particles are accelerated by shock waves 
created by colliding components of a jet. In such collisions, some fraction 
of the kinetic energy related to the bulk motion of the components is 
transformed into random kinetic energy of the particles.
This energy is then radiated away. The acceleration and evolution of the spectrum 
are usually described by the kinetic equation, which is a partial differential 
equation (e.g. \citealt{Kardashev62}). There are two general approaches 
to describe the acceleration in the equation. 

In the first approach, the acceleration term in the equation
describes the process inside a shock wave. This approach was used for
example by \citet{Kirk98}, where the evolution of the source was
represented by two kinetic equations. The first equation described the
acceleration inside the shock and the second the evolution of the
particle spectrum in the downstream region of the shock. Particles
were escaping into this region after the acceleration, creating a source 
for electromagnetic emission.  However, this is a rather complex 
approach and was applied only for the synchrotron emission.

The second approach is simpler.  There is no acceleration term,
i.e. no description of the acceleration process. Instead, an
``injection term'' defines directly the result of the acceleration.
One assumes a particle energy spectrum created by a hypothetical shock
and describes it with a few free parameters. This approach was used
for example in the model proposed by \citet{Chiaberge99}, where
different types of injection were investigated.

Since there is no perfect solution for modelling of variability in the
framework of the leptonic SSC scenario, we have chosen here the relatively
simple model proposed by \citet{Chiaberge99}. The model, as we will
show, provides very reasonable results while using only a minimum
number of free parameters. 

\section{Rapid variability}

High energy variability of TeV blazars has been observed many times by
orbiting X-ray experiments, as well as ground based gamma-ray
telescopes.  One of the most spectacular states of high activity was
observed in Mrk 501 in 1997.  The emission level and the spectrum of
the emission changed dramatically in a period of a few days
(\citealt{Catanese97}, \citealt{Pian98}, \citealt{Djannati99},
\citealt{Krawczynski00}).  Very rapid variability at time scales of
only a few hours was first observed simultaneously in X-rays and TeV
gamma rays in Mrk 421 and reported by \citet{Maraschi99}. Today Air Cherenkov
Telescopes like H.E.S.S. or MAGIC achieve major breakthroughs with
minute-scale observations in the VHE range. 

PKS 2155--304 has been intensively observed during the last years by
different instruments, showing variability at time scales from several
months to a few hours. For example \citet{Giommi98} observed X-ray
activity of this source with a variability time scale of a few
hours. Similar time scales of the X-ray activity were also reported by
\citet{Chiappetti99} and \citet{Edelson01}.  Observations made by
\citet{Tanihata01} show X-ray flux variations at time scales of about
one day. In the TeV range, the source has been detected by the
Durham Mk VI telescopes (\citealt{Chadwick99}).  This detection
remained unconfirmed until observations were made by the
H.E.S.S. experiment in 2002 and 2003 (\citealt{Aharonian05a}).
Variability on daily time scales was observed during that period.

Recently (28th of July 2006, i.e. MJD 53944), a huge TeV outburst
of PKS 2155--304 has been observed by the H.E.S.S. experiment
(\citealt{Aharonian07}).  The TeV emission level of the source changed
by a factor of about 20 within the observational period of about
1.5 h. Variability with a time scale of only a few minutes could be
observed, which presents the fastest variability ever seen in a
blazar. Unfortunately, there are no simultaneous X-ray observations
that could give us strong constraints on the emission process during this first huge outburst. After
this outburst, the source was being monitored by several experiments and especially almost
continuously by the UVOT, XRT and BAT experiments on board the {\it
  Swift} satellite (\citealt{Foschini07}) for a duration of several
days (from 29th of July to 22nd of August 2006, i.e. MJD 53945 to MJD
53969). The comparison of the {\it Swift} data with the observations
made by H.E.S.S. at that time (\citealt{Raue06}) shows a rather minor
change of the UV$\to$X-ray emission level (see Fig. 3 in
\citealt{Foschini07}) and a huge variation of the TeV
emission. Compared to most observations of blazars, this is a rather
unexpected behaviour; however, even more extreme events, so-called
orphan TeV flares, have already been observed in other sources
(\citealt{Krawczynski04b}, \citealt{Blazejowski05}). We will in the 
following try to explain this rapid TeV variability event in order 
to constrain the physical parameters of the source.
 
\section{The model}

Here we use the model proposed by \citealt{Chiaberge99}. They assume
that the source of VHE emission is created by a shock wave that
accelerates the electrons. The shock is perpendicular to the jet
symmetry axis and moves with constant relativistic velocity
($v'_s$). The particles accelerated by the shock up to extremely
relativistic energies escape into the downstream region of the shock,
where they generate most of the synchrotron and IC radiation. In
other words the shock is injecting relativistic particles into some
volume where most of the particle energy is dissipated by radiative
processes. Therefore, this volume is equivalent to the source volume.
For sake of simplicity, the cross section of the shock front
perpendicular to the jet symmetry axis is assumed to be square, of
scale $R$. The thickness of the shock front (parallel to the jet axis) 
is $r \ll R$. The shock is accelerating the particles and thus creating the 
source for a duration of $t_{\rm cr} = R/c$. Such an assumption eliminates the 
duration of the acceleration process as a free parameter and yields a final volume 
of the source of $R^3$.
The evolution of the source is also simulated
after the injection phase for a few crossing times up to almost
complete decay of the X-ray and the $\gamma$-ray emission.

The source volume is divided into 10$\times$10$\times$10 smaller cells
and calculations are performed in time steps of $\Delta t' = 0.1 R/c$.
This allows to take into account the external LCTE, which is the main
advantage of the model. In the first step, at time $t'=\Delta t'$, only
a narrow region just after the shock (at the distance from 0 to $v'_s
\Delta t'$) is filled up by particles. In the next step ($t'=2
\Delta t'$) the particles injected during the first phase are somewhat
``older" which means they have already lost some fraction of their
energy. They are located at a distance between $v'_s \Delta t'$ and $
2 v'_s \Delta t'$. It should be noted that, when solving the kinetic equation that
describes the evolution of the particle energy, we assume that the injection 
process works only during a time $t_{\rm inj} = 0.1 R / c$, necessary to create
a cell. Moreover injection is only active in the cells that are next to the shock 
front; for other cells we calculate only radiative cooling.
The shock is creating continuously new cells filled by ``fresh" particles, 
whereas old cells are moving systematically to larger distances. However, what
is observed in the comoving frame is significantly different from this
due to LCTE. The observed spectrum is produced by the electron
distribution at different stages of evolution. Therefore, to obtain
the total emission, it is necessary to sum up the different
contributions of different cells in an specific way, as described in
details in the original paper. 

The source is observed at an angle $\theta=1/\Gamma$ [rad], where
$\Gamma$ is the Lorentz factor that describes bulk motion of the
source. This means that in the source's comoving frame the emission is
observed at 90 degrees with respect to the jet symmetry axis and the
shock velocity vector. For such an angle the external LCTE has the
strongest impact on the evolution of the observed emission. Note that
for zero degree the external LCTE is negligible.

\citet{Chiaberge99} tested a few different types of injection spectra, here
we use only the very simple single power-law injection $Q(\gamma)=Q_0
\gamma^{-n}$ for $\gamma_{\rm min} \le \gamma \le \gamma_{\rm max}$,
where $\gamma$ is the Lorentz factor that describes the energy of the particle.
We neglect also a possible escape of the particles from the source. In
principle the particles may escape into regions of the jet where
the magnetic field strength is significantly lower in comparison to the
field inside the source. Therefore, synchrotron emission from such
regions should also be significantly lower and thus negligible.  However,
if the radiation field outside the source is high enough, the escaped
particles may still produce a significant amount of emission through
IC scattering. This is a quite complex scenario with additional free 
parameters not yet constrained by the data. Therefore, for sake of 
simplicity, we neglect here any possible escape.
Morover, we have also introduced some small improvements to 
the model. We apply a more precise description of the IC emissivity, which 
uses the Compton kernel computed by \citet{Jones68}. The kernel
describes scattering on an isotropic distribution of soft photons, considering
the full Klein-Nishina cross section in the head-on approximation (e.g.
\citealt{Inoue96}, \citealt{Sauge04}, \citealt{Moderski05}). The 
cooling rate ($\dot{\gamma}$) due to IC scattering is calculated more 
precisely than in the original model, using the Compton kernel mentioned 
above (e.g. \citealt{Sauge04}, \citealt{Moderski05}). We calculate the absorption 
of the TeV emission inside the source due to electron-positron pair production
using the approximation derived by \citet{Svensson87} (e.g. \citealt{Inoue96}).
The absorption is calculated separately inside each cell and 
absorption caused by the surrounding cells during the propagation of the emission 
from a given cell toward the observer is also taken into account. 
To correct the intrinsic TeV spectra for the IR extragalactic absorption, 
we use an optical depth derived by \citet{Stecker06} with the appropriate
correction (\citealt{Stecker07}), which was calculated in a 
$\Lambda$CDM universe for $h=0.7$, $\Omega_{\Lambda} = 0.7$ and $\Omega_m=0.3$.

There are several clear advantages to this model. The most significant one lies 
in the reduction of the number of free parameters needed to fully describe a single 
flaring event. For single power-law
injection we need in principle four free parameters: minimum and
maximum energy ($\gamma_{\rm{min}}, \gamma_{\rm{max}}$), density of
the injected particles ($Q_0$) and index of the power-law
($n$). However, precise values of $\gamma_{\rm min}$ and ${\gamma_{\rm max}}$ 
are not crucial for the model. To provide a significant
amount of low energy synchrotron photons for the IC scattering, it is
indeed sufficient to keep $\gamma_{\rm min}$ relatively small (e.g. $1 \le
\gamma_{\rm min} \le 10^3$). On the other hand, the value of
$\gamma_{\rm max}$ inside each cell at a given time is precisely
calculated according to the cooling conditions inside the
source. Therefore, it is sufficient to assume $\gamma_{\rm max}$ high
enough to be able to produce TeV gamma-rays. Simulating the emission
of six different sources we then use $\gamma_{\rm min} =1$
and $\gamma_{\rm max} = 10^6$. The index of the injected spectrum,
$n=2$, is the same in all our calculations. Three more free
parameters are required to completely describe a single source: the
source extension ($R$), the magnetic field strength inside the source
($B$), and the Doppler factor ($\delta$). 
This results in 7 free parameters, a number identical to the one required in 
the simple stationary scenario (see the introduction).
However, if this simple scenario uses a broken power-law particle
energy distribution (this is required in most cases) then 
the number of free parameters increases to 9. Moreover, it is not clear 
how such a broken power-law energy distribution could be created when 
for example the spectral index above the break is $>n+1$.  
On the other hand, the model we use requires only a single power-law 
injection to provide a self-consistent explanation for the observed 
X-ray spectra that one could approximate by broken power-law functions. 
Such broken synchrotron spectra are created by external LCTE. 
This is another advantage of the model and this shows also that the observed 
index of the synchrotron emission may have nothing to do with the index of 
the particle energy distribution inside the source. Since $\gamma_{\rm min}$, 
$\gamma_{\rm max}$ and $n$ have the same values in all our calculations and
minimal impact for the result of the modelling, we are left with only four
important parameters ($Q_0, B, R, \delta$) that describe the evolution
of a single source.  In comparison to the nine parameters required by
the simple stationary scenario, this is a quite simple description for a
single activity event. This is important when one has to explain
several flares one by one.

\begin{figure*}
\resizebox{\hsize}{14cm}{\includegraphics{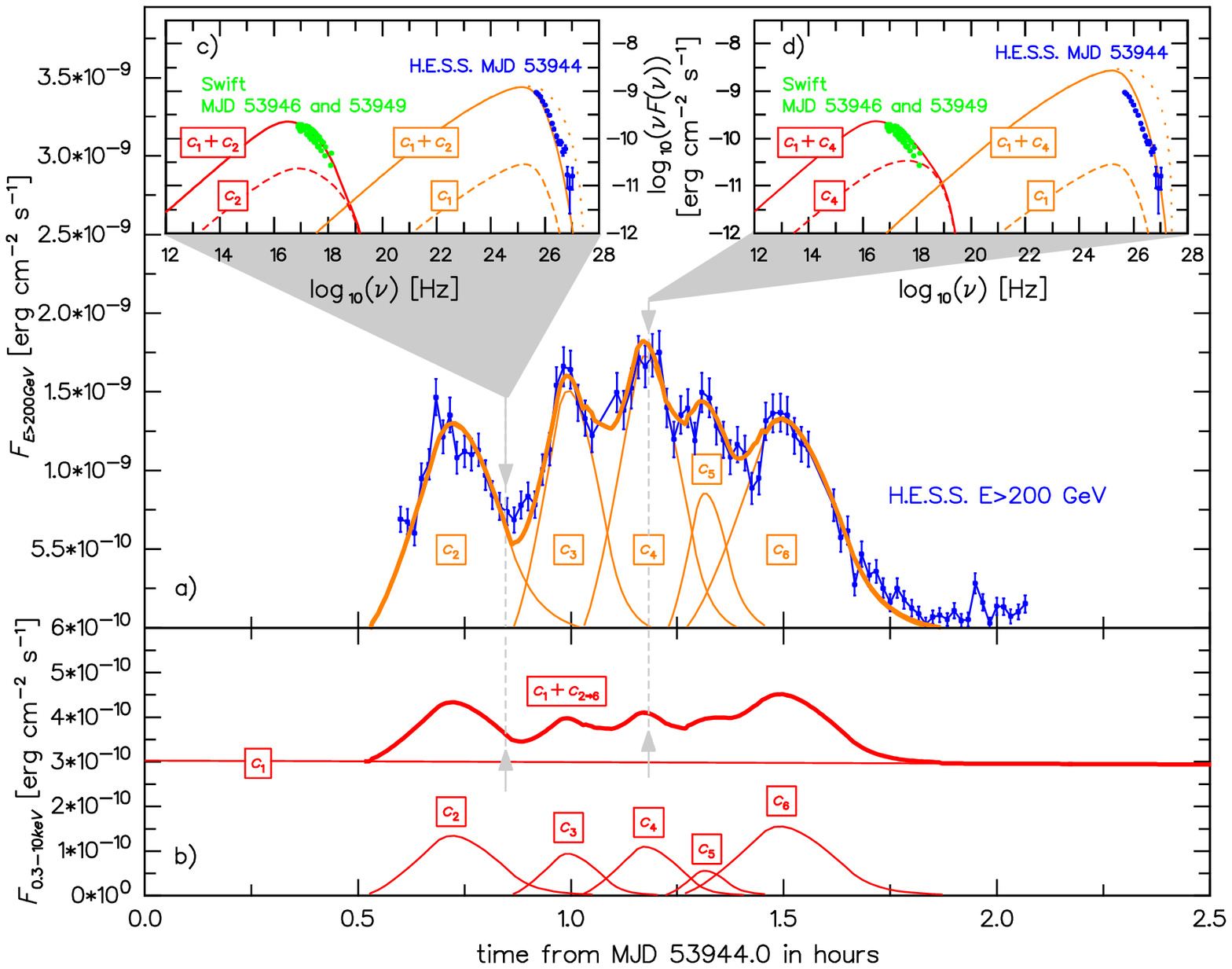}}
\caption{ Activity of PKS 2155--304 observed by H.E.S.S. (\citealt {Aharonian07}) \& {\it Swift}
          (\citealt{Foschini07}) in July and August 2006 and results  of our modelling.
          Panel a) shows the rapid variability observed by H.E.S.S. and the calculated light curve
          (thick solid line), where the contributions from simulated jet components ($c_{2\to6}$)
          are indicated by thin solid lines. A theoretical light curve has also been calculated
          in the X-ray range (panel b). Here, an extended, slowly evolving jet component ($c_1$)
          dominates the emission over the other jet components. Sub-panels c) and d) show the
          observed average spectrum in the TeV range and two spectra obtained by Swift a few
          days later. In the sub-panels, we also show spectra calculated for two
          arbitrarily chosen times (MJD 53944.035 and 53944.049) corresponding to
          medium (c) and high (d) TeV emission levels. The solid
          lines in the sub-panels show the total emission; the dashed
          lines indicate contributions from individual components. One
          can see that the background component ($c_1$) dominates in the
          X-ray range, whereas the foreground components (here $c_2$ and
          $c_4$) dominate the emission in the TeV range. Calculated not 
          absorbed intrinsic TeV spectra are indicated by doted lines.
         }
\label{fig_lum}
\end{figure*}

The model does not describe internal LCTE, which is the main drawback
of this scenario. However, this effect is not very important in the case
where single power-law particle energy distribution is injected. After
the injection, the particle energy spectrum inside the cells is
modified only by radiative cooling that is systematically reducing 
{$\gamma_{\rm max}$}. This means that the synchrotron emission produced by 
the most energetic particles is also systematically reduced. However, 
this emission contributes little to the IC scattering due to the 
Klein-Nishina restriction. On the other hand, the synchrotron emission 
produced by the particles with medium and low energies remains 
unaffected by the cooling for a relatively long time. Thus the 
synchrotron emission produced by those particles remains equal 
and constant inside most of the cells and is largely responsible
for the IC scattering.  

\section{Results}

There are no X-ray observations simultaneous with the rapid
variability observed by H.E.S.S. on the 28th of July 2006. Therefore,
to constrain the model we decided to use the data obtained by {\it Swift}
during the several day long campaign of observations that
was triggered by this rapid TeV activity. The X-ray light curve and
spectrum does not show a significant change of the emission level,
whereas TeV observations (\citealt{Raue06}) shows a
dramatic change of the TeV emission level during this campaign. Thus, 
we decided to simulate an event where X-ray emission remains
almost constant during the TeV flaring activity. As we already 
mentioned such activity was already observed at least two times 
in other TeV sources.

\begin{table}
\caption{The most important parameters used for the modeling, where $c_1$ 
        indicates background component of the jet and $c_{2 \to 6}$ indicate 
        parameters that describe foreground components. The other parameters: 
        $n=2$, $\gamma_{\rm min} = 1$ and $\gamma_{\rm max} = 10^6$ are the same 
        for each component. Note that we specify particle density inside source 
        ($t_{\rm inj} Q_0$) instead of the injection rate ($Q_0$) to give 
        possibility for easy comparison between our calculations and results 
        that may give simple stationary SSC scenario.}
\label{tab_params}        
\begin{center}
\begin{tabular}{lllllllr}
\hline
 name & $c_1$ & $c_2$ & $c_3$ & $c_4$ & $c_5$ & $c_6$ & unit\\
\hline
$\delta$ &  20        & 30  & 30  & 30 & 30 & 30 & \\
$ R$     & 3 10$^{3}$ & 5.2 & 3.5 & 4  & 2.3 & 6 & 10$^{14}$ cm\\ 
$ B$     & 0.02 & 0.1 & 0.1 & 0.1 & 0.1 & 0.1 & G\\
$ t_{\rm inj} Q_0$ & 8 10$^{-6}$ & 3.2 & 7.95 & 6.8 & 12 & 2.45 & 10$^{7}$ cm$^{-3}$ \\
\hline
\end{tabular}
\end{center}
\end{table}

However, such a situation  is problematic for SSC modelling, where
the synchrotron emission that provides low energy photons for the IC
scattering should usually vary as much as the IC emission. Only in the 
quite unrealistic case where only the particle density inside the 
source is varying may the IC emission increase or decrease as
the square of the synchrotron emission ($F_{\rm IC}(\nu, t) \propto \left[ 
F_{\rm synch}(\nu, t) \right]^2$).
However, this anyway may be diluted by the LCTE (\citealt {Katarzynski05}).
The simple solution for this problem is to assume that the X-ray emission is
dominated by a relatively large component from a jet that provides almost
constant emission on a time scale of several days. Since the density of such
a large component is relatively low, the level of the TeV emission must also
be negligibly low. The existence of such a component is strongly supported
by the observations made by {\it Swift}, where for several days a nearly
constant emission level was observed. The H.E.S.S. observations
show that, in addition, relatively small components can appear inside the jet.
These components may not be strong enough to be dominant in X-rays, but
may be very compact and therefore very strong in the TeV range.

We simulate in our model one large jet component (we call it background
component because it is not visible in the TeV range), and five small
components (here called foreground components) dominant at TeV energies.
The main difference between the background component and the foreground
components appears in the estimate of the size and the particle  
density. Moreover, the small components are likely located at a distance of less than 1  
pc from the center, whereas the large component should be located further downstream
in the jet, a few parsec from the center.

In Fig.~1 we show the light curve and average spectrum obtained by
H.E.S.S.  on MJD 53944 as well as the spectra obtained by $Swift$ a
few days later.  We also show in this figure the results of our
calculations, where rapid TeV variability is well reproduced by the
superposition of the IC emission of the five foreground jet components
(denoted $c_{2 \to 6}$).  Note that the IC emission of the background
component ($c_1$) is negligible in the TeV energy range. On the other
hand, this component dominates the emission in the X-ray range. We show
a theoretical light curve calculated in the $Swift$ energy range
0.3--10 keV, where emission from the background component appears to be
almost constant in the relatively short time interval (2.5 h)
presented in the figure. The total duration of the background
component activity, which is calculated in the same way as the activity
of the foreground components, is about 15 days. This long
term activity starts 10 days earlier than the first foreground
component ($c_2$) flare. Finally, we show synchrotron and IC spectra
calculated for two arbitrarily selected times (MJD 53944.035 and
53944.049) that correspond to medium and high TeV emission
levels. Note that the calculated spectra are not well constrained,
since we have only an average observed TeV spectrum and no
simultaneous X-ray observations. Detailed values
of the physical parameters used for the calculations are given in Tab.~1. 
Note that in order to precisely explain the TeV light curve we had
to modify only the sizes of the small sources that is constrained by 
the observed variability time scales and the particle injection rate 
to explain correctly the level of the emission. 

Rapid variability requires a relatively small and dense source, which
may be optically thick for TeV emission due to pair production inside 
the source. The simplest solution for this problem is to assume a relatively 
large value of the source Doppler factor ($\delta \gtrsim 50$), which decreases
the observed variability time scale and increases the observed emission level.
However, such a large value of the Doppler factor might also significantly increase
the scattering of electrons inside the source on the external ambient radiation field. 
Therefore, the external IC scattering might become the dominant 
process that produces most of the TeV emission. Such a scenario was recently
proposed by \citet{Begelman07}. However, the efficiency of the external 
IC scattering depends on the intensity of the radiation field and 
also on the distance from the center where the source is created.
It may be possible that the intensity is too low and the distance is
too large for this process to be important, as has usually been assumed 
in the modeling of the TeV blazar activity up to now.

An alternate scenario to the external IC scattering was proposed recently
by \citet{Ghisellini08}. They assume that a fast jet ($\delta \lesssim 30$)
contains even faster moving compact sources ($\delta \gtrsim 50$). The radiation
field produced by the jet is amplified in the comovig frame of the compact 
sources due to the difference in velocities, which increases the IC emission 
produced by these sources. However, measurements of the motion of parsec-scale jet 
components in PKS~2155--304 at radio frequencies suggest a value
of the Doppler factor of a few (\citealt{Piner04}). 

We have carefully calculated the absorption due to pair production 
inside the source using a moderate value of the Doppler factor ($\delta=30$). 
We found this process to be negligible in our approach.  There are two reasons 
for this. First, the variability time scale that we are trying to explain 
is still relatively long (10-15 minutes), 
and second, more importantly, in all our 
calculations of the emission from small sources, the synchrotron emission 
level is significantly lower than the IC radiation level. It should be noted 
that we have also verified that by trying to explain a variability time scale 
of about 5 minutes, as well as by trying to keep $\nu F_{\rm syn, X-ray} \simeq 
\nu F_{\rm IC, TeV}$ that is frequently observed in TeV blazars, we may 
obtain significant absorption of the TeV emission. Our results show that the 
classical SSC approach used frequently to explain emission of TeV blazars can 
also explain rapid TeV variability if we assume that the X-ray emission is 
dominated by an extended, slowly evolving source.

\section{Conclusions}

We have shown in this paper that the rapid TeV variability of PKS 2155--304 
can be well explained using a standard SSC approach while taking into account 
the particle evolution and the external LCTE. In our approach the internal
LCTE is of minor importance. The model we use requires in principle four 
free parameters to describe a single activity event. Therefore, we may compare 
the physical parameters derived from such modelling with the parameters
derived form previous estimations of blazar emission (e.g. \citealt{Tavecchio98}). 
This shows that our parameters are similar to other results (e.g., $B \leq 0.1$ G 
and $\delta = 20 \to 30$); only the size ($R \simeq 5 \times 10^{14}$ cm) of the 
small sources is about one order of magnitude smaller than the 
values estimated so far.

\section*{Acknowledgments}

We thank the anonymous referee for a number of constructive comments that 
greatly helped us to improve the paper. This work was partially supported 
by the PAN/CNRS LEA Astro-PF.

\label{lastpage}

\end{document}